\documentclass[twocolumn,preprintnumbers,nofootinbib,prd]{revtex4}

\usepackage{graphicx}

\usepackage[hypertex]{hyperref}
\usepackage{amsmath, amssymb}

\newcommand{\lsim}{\lesssim}
\newcommand{\ord}[1]{\mathcal{O}{(#1)}}

\newcommand{\gsim}{\gtrsim}
\newcommand{\beq}{\begin{equation}}
\newcommand{\eeq}{\end{equation}}

\begin{document}

\pagestyle{plain}

\preprint{BNL-HET-07/8}

\title{Constraining Unparticle Physics with Cosmology and Astrophysics}

\author{Hooman Davoudiasl\footnote{email: hooman@bnl.gov} }

\affiliation{Department of Physics, Brookhaven National Laboratory,
Upton, NY 11973, USA}


\begin{abstract}

It has recently been suggested that a scale invariant ``unparticle" sector
with a non-trivial infrared fixed point may couple to the Standard Model (SM)
via higher dimensional operators.  The weakness of such interactions hides the
the unparticle phenomena at low energies.  We
demonstrate how
cosmology and astrophysics can place
significant bounds on the strength of
unparticle-SM interactions.  We also discuss the
possibility of a having a non-negligible
unparticle relic density today.

\end{abstract}
\maketitle


In a recent paper \cite{Georgi:2007ek}, Georgi has suggested that a
scale-invariant sector with a non-trivial infrared fixed point may
couple to the Standard Model (SM) via higher-dimensional operators
cutoff by a large scale. Due to its scale-invariance, this sector is
not described in terms of particles.  Thus, the corresponding
phenomenology would be different from all other extension of the SM,
like supersymmetry and extra dimensions, which are based on a
particle interpretation.   Following Ref.~\cite{Georgi:2007ek}, we
will refer to this sector collectively as ``the
unparticle". Subsequent works in Ref.~\cite{Georgi:2007si} and in
Refs.~\cite{Cheung:2007ue,Luo:2007bq,Chen:2007vv,Ding:2007bm,
Liao:2007bx,Aliev:2007qw,Li:2007by,Duraisamy:2007aw,Lu:2007mx} have
presented some of the interesting collider and flavor phenomenology
of the unparticle that may appear above the TeV scale or in
precision measurements.  Refs.~\cite{Stephanov:2007ry,Fox:2007sy}
elucidate some of the more theoretic aspects of unparticle physics,
in the context of AdS$_5$ deconstruction and supersymmetry, respectively.

In this Letter, we will examine possible effects of the unparticle sector on cosmology
and astrophysics.  We will use considerations based on preserving the success of Big Bang
Nucleosynthesis (BBN) and stellar evolution to place constraints on this new physics.
Similar cosmological and astrophysical constraints have been considered
in Ref.~\cite{Gubser:1999vj},
in the specific context of AdS/CFT correspondence \cite{Maldacena:1997re}.
Here, we also discuss a possible mechanism for generating a significant relic
unparticle density that could survive until today and contribute to the cosmic dark
density (we cannot call this ``dark matter").
Before presenting our analysis, we will first review the basic
framework outlined in Ref.~\cite{Georgi:2007ek}.

The main assumption here is that, at high energy, the SM and the
fields of a Banks-Zaks (BZ) theory \cite{Banks:1981nn}, with a
non-trivial infrared fixed point, interact via the exchange of
particles of mass $M_U$:
\beq {\cal L}_{BZ} = \frac{O_{SM}\,
O_{BZ}}{M_U^k},
\label{smbz}
\eeq
where $O_{SM}$ is an SM operator
of mass dimension $d_{SM}$ and $O_{BZ}$ is a BZ operator of mass
dimension $d_{BZ}$.  In Eq.(\ref{smbz}), we have taken the
coefficient of the operator to be unity. Upon the onset of
scale-invariance, the interactions of the BZ fields give rise to
dimensional transmutation at a scale $\Lambda_U$, below which ${\cal
L}_{BZ} \to {\cal L}_U$, where
\beq {\cal L}_U =
C_U\frac{\Lambda_U^{d_{BZ}-d_U}}{M_U^k} O_{SM}\, O_U.
\label{smu}
\eeq
Here, $C_U$ is a coefficient in the low energy effective theory
and $O_U$ is an unparticle operator of dimension $d_U$.  
Generally speaking, each unparticle operator
has a different coefficient.  To avoid complicating the notation, we
will use $C_U$ to denote all such coefficients, with the
understanding that they are not assumed to be universal.

It was shown in Ref.~\cite{Georgi:2007ek} that the phase space $d\Phi$ for
an unparticle operator of dimension $d_U$ is the same as the phase
space for $n=d_U$ massless {\it invisible} particles.  This is an
interesting and exotic feature of this sector, since $d_U$ is not
necessarily integral.  Thus, $d\Phi(d_U)$
is proportional to the coefficient function
\beq
A_{d_U}
= \frac{16 \pi^{5/2}}{(2\pi)^{2 d_U}}
\frac{\Gamma(d_U+1/2)}{\Gamma(d_U-1)\Gamma(2d_U)}.
\label{AdU}
\eeq
In what follows, we will not present exact expressions, as they would
be rather unwarranted at the level cosmological and astrophysical
effects are treated here. The $A_{d_U}$ phase
space factors for $d_U\sim 1$ will not change our conclusions
significantly, and are hence ignored.

{\bf Unparticle Cosmology:}  In order to have a handle on unparticle cosmology,
we begin with the unparticle equation of state.  The type of
substance that makes up the unparticle sector can only be described
by a massless equation of state.  In fact, it is well-known that
pure radiation is classically scale invariant, $T_\mu^{(Rad)\mu} =
0$, and only develops scale-dependence through quantum mechanical
interactions.  In the case of the unparticle, the scale
invariance persists even at the quantum level and hence we adopt the
trivial equation of state
$
p_U = \rho_U/3
$
for this sector \cite{Gubser:1999vj}, where $\rho_U$ is the energy density
and $p_U$ is the pressure of the unparticle.

Given the success of BBN in predicting light element abundances, we
are compelled to make sure the unparticle will not significantly
change the physics of this epoch. For a general scale invariant sector
$\rho_U \sim T_U^4$ \cite{Gubser:1999vj}, up to an unknown
coefficient that we take to be ${\cal O}(1)$, where $T_U$ is the
unparticle temperature. Hence, one way to ensure that $\rho_U$ does
not interfere with BBN, is to require $T_U \ll T$, where $T \sim
1$~MeV is the temperature of the SM radiation during this epoch.  To
end up with a cold unparticle sector, we may assume that $O_U$
decoupled from the SM at an earlier time and did not get reheated in
the subsequent evolution of the radiation dominated universe.
However, we will see that, for certain ranges of parameters, the
unparticle recouples at lower temperatures.  For such cases, we must
demand that the unparticle stays decoupled throughout BBN.

In order to get a quantitative estimate,
we consider the case $O_{SM}= {\bar \psi}
\gamma^\mu \psi$, with $\psi$ an SM fermion.
Since we will mostly rely on
dimensional analysis for our estimates, this case
well-represents other dimension-3
SM operators.  We note that the lowest dimension
gauge-invariant SM operator we can right down
is $\phi^\dagger \phi$, where $\phi$
is the {\it elementary} Higgs field \cite{Fox:2007sy};
an interesting analysis of this
coupling has been provided in Ref.~\cite{Fox:2007sy}.
Here, we will study the consequences of the scenario in
Ref.~\cite{Georgi:2007ek}, with scale invariance.
We thus ignore Higgs effects which could be suppressed
due to, say, compositeness, above the electroweak scale.
Then, SM dimension-3
operators are the most important ones for our
analysis and we will focus on them.

To study how the unparticle decouples, we consider
\beq
{\cal L_\psi} = C_U\frac{\Lambda_U^{d_{BZ}-d_U}}{M_U^k}
{\bar \psi} \gamma_\mu \psi \, O_U^\mu,
\label{sm3u}
\eeq
with $d_{BZ}=k+1$.
The unparticle-SM interactions will drop out of equilibrium
once the rate $\Gamma_\psi$ falls below $H$, the relevant
Hubble constant.  At temperature $T$, we have
\beq
\Gamma_\psi \sim
\left|\frac{C_U \Lambda_U^{k+1-d_U}}{M_U^k}\right|^2
T^{2 d_U - 1}.
\label{Gampsi}
\eeq
During the radiation domination era,
$H\sim T^2/M_P$, where $M_P \sim 10^{19}$~GeV is the Planck scale.
Requiring $\Gamma_\psi \lsim H$ then gives
\beq
\left|\frac{C_U \Lambda_U^{k+1-d_U}}{M_U^k}\right|
T^{d_U - 1} \lsim (T/M_P)^{1/2}.
\label{decoupling}
\eeq

It is reasonable to assume $k=2$, if massive boson
in the ultraviolet couple the BZ and SM sectors.  These particles
must be heavier than $\sim 1$~TeV, since we have no evidence
for them.  In fact, they are likely much heavier
if one considers precision data,
leading to $M_U \sim 10^3$~TeV as the cutoff scale.
We will present most of our results for cases $d_U=1,3/2,2$.
Here, we only consider a new scale-invariant sector.  
Without the assumption of full conformal symmetry, our
analysis is not subject to spin-dependent constraints
which would otherwise apply to $d_U$ \cite{Mack:1975je}.
Ref.~\cite{Georgi:2007si} only considers non-integer
$d_U$ to avoid possible pathologies.  Since most of our
numerical results are given as orders of magnitude, we may
take our results for integer $d_U$ to be
also the relevant estimates for sufficiently close
non-integer values.

Eq.~(\ref{decoupling}), implies that for $1\leq d_U\leq 3/2$, the
unparticle rate of thermal interactions redshifts more slowly than $H$.
For this range of values, we must then ensure that the
unparticle sector remains decoupled throughout BBN.
We hence demand $\Gamma_\psi \lsim H$ for $T\sim 1$~MeV. 
With $M_U\sim 10^3$~TeV, we find 
${\tilde \Lambda}\lsim 3,\, 100 ~{\rm GeV}$, 
for $d_U=1, 3/2$, respectively, where ${\tilde \Lambda} \equiv
C_U^{1/(k+1-d_U)}\Lambda_U$.

\begin{figure}
\includegraphics[width=0.48\textwidth]{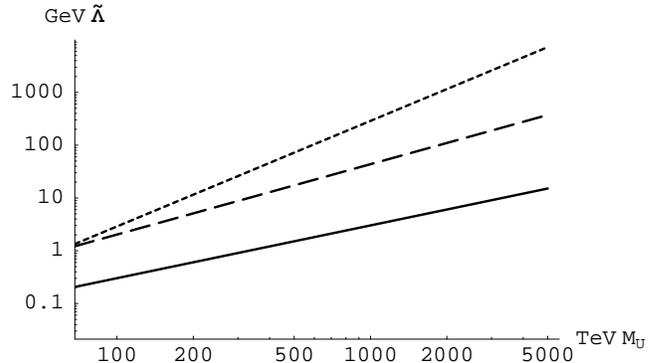}
\caption{BBN constraints on the unparticle sector
from decoupling requirements.  
We have used Eq.~(\ref{decoupling}), with $k=2$.
From bottom to top, the lines are for $d_U = 1,3/2,2$,
respectively.  The excluded region
lies above the lines.}
\label{fig1}
\end{figure}

For $d_U > 3/2$, the rate $\Gamma_\psi$ redshifts faster than $H$
and we have a decoupling behavior.  In this case,
we require that the unparticle decouple before BBN,
but not get reheated after SM phase transitions.
A minimal assumption is that this decoupling happened before the
Quantum Chromo-Dynamics (QCD)
phase transition at $T \sim 1$~GeV.  The released latent heat after
the transition only heats up the SM radiation.  This leaves the
unparticle colder, resulting in $\rho_U \ll \rho_{SM}$ during BBN.
We note that this conclusion holds, as long as the unparticle sector
does not have a very large number of degrees of freedom
\cite{Gubser:1999vj}.  For $T \sim 1$~GeV
and $M_U\sim 10^3$~TeV, we find:
${\tilde \Lambda}\lsim 100~{\rm GeV}$ with $d_U=2$.
We have plotted the above BBN bounds on ${\tilde \Lambda}$ 
for a range of $M_U$ values and $d_U=1,3/2,2$,
in Fig.~\ref{fig1}.
Stronger constraints apply if QCD phase transition
does not heat up the visible sector enough to marginalize the unparticle
contribution.

Having a decoupled or colder unparticle gas
is not necessarily enough to ensure that
our standard picture of cosmology is intact.  In fact, we will have to consider
the possibility that the SM plasma can lose energy into
the unparticle and undergo evaporative cooling.  This constraint has been
previously studied in the context of models with large extra dimensions
\cite{Arkani-Hamed:1998rs} and AdS/CFT correspondence \cite{Gubser:1999vj}.
By dimensional analysis, ensuring that evaporative cooling
is sub-dominant to the one from
the expansion of the universe yields the minimal condition
$\rho_{SM} H \gsim \Gamma_\psi T^4$,
where the left-hand side is the cooling rate due to the expansion and
the right-hand side is given by evaporation into the unparticle.
To be specific, we will take the BBN era with
$T\sim 1$~MeV and require that the
cooling during this period is dominantly from cosmological expansion.
We find that this does not yield stronger bounds than decoupling;
with $M_U\sim 10^3$~TeV we get
${\tilde \Lambda}\lsim 3,\, 100,\, 
10^4~{\rm GeV}$ for
$d_U=1,3/2, 2$, respectively.
For $d_U = 3/2$ the
bound becomes $T$-independent.  We have assumed that
BBN is characterized by one temperature of $\ord{\rm MeV}$.
This is a simplification,
however changing $T$ by factors of order
unity does not change our conclusions
significantly.  For $d_U=2$, cooling
sets a weaker upper bound on $\Lambda_U$ than before.

Here, we note that if the unparticle recouples after BBN,
as in the above discussion for $1\leq d_U \leq 3/2$, it could
come into equilibrium with neutrinos, with $\psi=\nu$ in
Eq.~(\ref{sm3u}).  In this case, neutrinos will cease to free-stream
and a $\nu$-$U$ fluid gets established.  This can lead to non-standard
shifts in the location of the acoustic peaks of the cosmic microwave
background \cite{Chacko:2003dt}.  The cosmic evolution
of the $\nu$-$U$ fluid can have
interesting signatures that merit more consideration.  However,
these will be outside the scope of the present work.

Given the above discussion, thermally produced unparticle
will not be a significant component of the ``invisible"
energy density today.  To see this, note that the
unparticle redshifts like
radiation and that it is likely no hotter than the relic photon
temperature of order $10^{-4}$~eV.  However,
this conclusion can change if the unparticle is produced
non-thermally.  This can happen if the unparticle sector
is coupled to ``dark matter operators".  If the dark matter
produced in the early universe
is unstable and can decay into other dark matter plus the
unparticle $U$, then we can expect to have a more sizable
$U$-density today.  Let us consider the following interaction
between dark matter and the unparticle
\beq
{\cal L}_{DM} = C_U\frac{\Lambda_U^{d_{BZ}-d_U}}{M_U^k}
O_{DM}\, O_U,
\label{dmu}
\eeq
where $O_{DM}$ is made out of dark matter fields.  Assuming
a dimension-3 fermionic dark matter operator
${\bar \chi_1}\gamma^\mu \chi_2$, with
$m_{\chi_1} \gsim m_{\chi_2} \sim 10^2$~GeV, as expected for
WIMP dark matter, the
decay rate $\Gamma_\chi$ for $\chi_1 \to \chi_2 U$ is given by
\beq
\Gamma_\chi \sim
\left|\frac{C_U \Lambda_U^{k+1-d_U}}{M_U^k}\right|^2
m_{\chi_1}^{2 d_U - 1}.
\eeq
Let us first consider $k=2$, $d_U=2$; we find
$\Gamma_\chi \sim  10^{-15} |C_U \Lambda_U|^2$~TeV$^{-1}$.  For the
above decay to take place in the recent cosmological epoch, to avoid
red-shifting the unparticle away, we require $\Gamma \lsim H_0$, where
$H_0 \sim 10^{-33}$~eV.  We then find $|C_U \Lambda_U| \lsim 10^{-3}$~eV.
However, a value of $\Lambda_U$ close to this limit is rather inconsistent
with our assumption that dimensional transmutation in the BZ sector takes
place above the energy scales we are considering.
Another motivated mass scale above the weak scale is $M_{GUT}\sim 10^{15}$~GeV.
If we choose $M_U \sim M_{GUT}$, we get $|C_U \Lambda_U| \lsim 10^{3}$~TeV.

Current precision for the measured cosmological
parameters \cite{Spergel:2006hy}
allow one dark matter component, comprising roughly
$5-10\%$ of the original
WIMP population, to decay into the unparticle.
We may then expect that today the
energy densities in baryonic matter and the
unparticle are roughly of the same order.
If the unparticle thermalizes by the present time, we may
expect it to have a temperature $T_0^U$ roughly given by
\beq
x \Omega_{DM} \sim (T_0^U)^4,
\label{T0U}
\eeq
where $x$ is the small fraction of today's
dark matter density $\Omega_{DM} \sim (1.5\times 10^{-3}~{\rm eV}^4)$
that decayed recently.  In Eq.~(\ref{T0U}), we have ignored effects
coming from different redshifts of
matter and radiation. We may then expect
an unparticle gas of temperature
$T_0^U \sim 10^{-3}$~eV for $x\sim 0.1$.

For this substance to be a viable component of cosmic energy
density today, we must consider whether it can decay
back into the SM.  A reasonable assumption is that
such a cold scale invariant gas can return back into the visible sector only
by transferring its energy into massless photons.  We then
consider the interaction
\beq
{\cal L}_{\gamma} = C_U\frac{\Lambda_U^{k-d_U}}{M_U^k}
F_{\mu\nu}F^{\mu\nu}\, O_U,
\label{gam}
\eeq
where $F_{\mu\nu}$ is the photon field strength tensor.
We estimate the rate $\Gamma_\gamma$
of energy leakage from the
unparticle into photons by
\beq
\Gamma_\gamma \sim |C_U|^2 \left(\frac{\Lambda_U}{M_U}\right)^{2k}
\left(\frac{T_0^U}{\Lambda_U}\right)^{2 d_U}T_0^U.
\label{Gamgam}
\eeq
If this leakage occurs on time scales short compared to Hubble time 
it can distort the cosmic background radiation. We thus require  
$\Gamma_\gamma \ll H_0$.  Choosing $k=2$ and $M_U\sim 10^3$~TeV again, yields 
$\Lambda_U\ll |C_U|^{-1}10^6~{\rm TeV}$ for $d_U=1$, 
implying that we do not have a constraint. For $d_U=2$ the bound is also
well-satisfied.

{\bf Unparticle Astrophysics:}  New physics that includes very light
degrees of freedom can be strongly constrained
by astrophysical processes.  Examples of such physics are
axions \cite{Yao:2006px} and light graviton Kaluza-Klein modes
\cite{Arkani-Hamed:1998rs,Cullen:1999hc}.  It is then
interesting to inquire how the
unparticle interactions can be constrained by these processes.
One such bound can be obtained by considering the SN 1987A.
The observation of this supernova constrains the emission of
non-neutrino species from its hot core, with $T_{SN}\sim 30$~MeV.
The bound on the axion coupling constant
$f_a \gsim 10^9$~GeV \cite{Yao:2006px} can be translated into a bound on
unparticle-nucleon interactions as follows.

Let us consider the interaction
\beq
{\cal L_N} = C_U\frac{\Lambda_U^{k+1-d_U}}{M_U^k}
{\bar N} \gamma_\mu N \, O_U^\mu,
\label{NNu}
\eeq
where $N$ is a nucleon.  The coupling of the axion to the nucleon
tends to zero in the non-relativistic limit.  Therefore, the
relevant dimensionless effective coupling for the axion is given by
\beq
g_a^{SN} \sim (T_{SN}/f_a) \sim 3\times 10^{-11}.
\label{ga}
\eeq
To avoid over-cooling
the supernova via unparticle emission, we then require, for
$k=2$,
\beq
\left|\frac{C_U \Lambda^{3-d_U}}{M_U^2}\right| T_{SN}^{d_U-1}
\lsim g_a^{SN}.
\label{SN}
\eeq
For $M_U\sim 10^3$~TeV,
${\tilde \Lambda}\lsim 5,\,30,\,10^3~{\rm GeV}$ 
with $d_U = 1, 3/2, 2$.
There is also a very stringent bound on axion-photon
coupling from the evolution of globular clusters;
$g_a^{GB} \lsim 10^{-10}$~GeV$^{-1}$ \cite{Yao:2006px}.
The relevant temperature here is $T_{GB} \sim 10$~keV and hence
$g_a^{GB}T_{GB} \lsim 10^{-15}$.  Using the interaction
in Eq.~(\ref{gam}), with $k=2$, $d_U=1$, and $M\sim 10^3$~TeV,
we get $\Lambda_U \lsim |C_U|^{-1} 100$~GeV, whereas
for $d_U=3/2$ we get $\Lambda_U \lsim |C_U|^{-1} 10^7$~TeV.
The case with $d_U=2$ does not yield a new limit.

\begin{figure}
\includegraphics[width=0.48\textwidth]{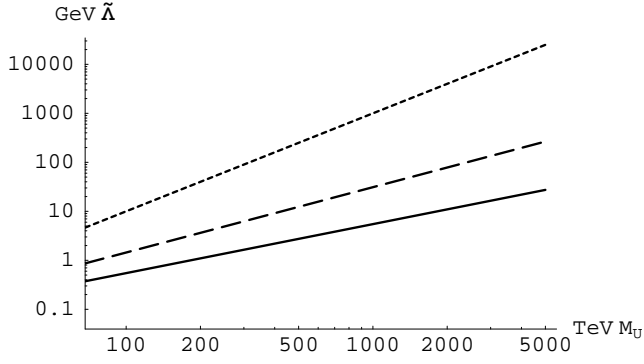}
\caption{Supernova over-cooling constraints on the unparticle-nucleon
interactions, from Eq.~(\ref{SN}).  Other conventions are as in Fig.~\ref{fig1}.
}
\label{fig2}
\end{figure}

Before closing, it is interesting to see how the above considerations
can affect the collider and precision phenomenology of unparticle physics.
For example, let us take the results of Ref.~\cite{Cheung:2007ue} for
Drell-Yan processes and $(g-2)_\ell$.  A typical range of $d_U$ in
Ref.~\cite{Cheung:2007ue} is $3/2\leq d_U \leq 2$ and $\Lambda_U = 1$~TeV
has been assumed for the above processes.  To make contact with their notation,
we define $\lambda_1 \equiv C_U (\Lambda_U/M_U)^k$.  Then,
our Eq.~(\ref{decoupling}) becomes
$\lambda_1 (T/\Lambda_U)^{d_U-1} \lsim (T/M_P)^{1/2}$. For
$d_U=3/2$ we get $\lambda_1 \lsim 10^{-16}$, where as
for $d_U=2$ and $T\sim 1$~MeV we get $\lambda_1 \lsim 10^{-5}$, whereas 
in Ref.~\cite{Cheung:2007ue},
$\lambda_1 \geq 10^{-3}$.  Hence, cosmological constraints severely affect the
viable parameter space relevant for these processes.

In summary, we considered how current standard cosmology
and astrophysics place bounds on unparticle interactions.
The suppression power of the cutoff $M_U$
scale was taken to be $k=2$, as would
often be the case for higher dimensional operators.
The strongest bounds we obtained are for $d_U=1$,
imposed by the success of BBN and agreement with the SN 1987A data.  We also
considered a scenario in which couplings of a WIMP-type dark matter to
the unparticle lead to a present ``dark-unparticle''
energy density at a level near that of baryons.
The bounds in our work can be useful guides for
unparticle model-building and phenomenology,
as demonstrated for some of the hitherto
studied collider and precision phenomenology of unparticle physics.

\acknowledgments

The author would like to thank Mark Wise for discussions and
Yu Nakayama for comments on the results of
Ref.~\cite{Mack:1975je} in relation to unparticle phenomena.
This work was supported in part by the United States Department of
Energy under Grant Contracts DE-AC02-98CH10886.

\end{document}